\documentclass[aps,prb,floatfix,twocolumn,superscriptaddress,showpacs,amsmath,amssymb]{revtex4-1}
\usepackage{graphicx}
\usepackage{epsfig} 
\usepackage{amssymb}
\usepackage{amsmath}
\usepackage{color}
\allowdisplaybreaks

\begin{document}
\title{Correlation energy within exact-exchange ACFD theory: systematic \\development and simple approximations} 

\author{Nicola Colonna}
\affiliation{International School for Advanced Studies (SISSA), Via Bonomea 265, I-34136 Trieste, Italy} 
\author{Maria Hellgren}
\affiliation{International School for Advanced Studies (SISSA), Via Bonomea 265, I-34136 Trieste, Italy} 
\author{Stefano de Gironcoli}
\affiliation{International School for Advanced Studies (SISSA), Via Bonomea 265, I-34136 Trieste, Italy}
 \affiliation{CRS Democritos, CNR-IOM Democritos, Via Bonomea 265, I-34136 Trieste, Italy}

\date{\today} 

%\pacs{71.10.Fd, 05.30.Fk, 05.70.Ln}

\begin{abstract}

We have calculated the correlation energy of the homogeneous electron gas (HEG) and the dissociation energy curves 
of molecules with covalent bonds from a novel implementation of the adiabatic connection fluctuation
dissipation (ACFD) expression including the exact exchange (EXX) kernel. The EXX kernel
is defined from first order perturbation theory and used in the Dyson equation of
time-dependent density functional theory. Within this approximation (RPAx), the correlation energies
of the HEG are significantly improved with respect to the RPA up to densities of the order of
$r_s \approx 10$. However, beyond this value, the RPAx response function exhibits an unphysical divergence
and the approximation breaks down. Total energies of molecules at equilibrium are also highly accurate but we find a similar 
instability at stretched geometries. Staying within an exact first order approximation to the response function 
we use an alternative resummation of the higher order terms. This slight 
redefinition of RPAx fixes the instability in total energy calculations without compromising the 
overall accuracy of the approach.
\end{abstract}
\maketitle

\section{Introduction}\label{sect:sect0}

Kohn-Sham (KS) methods that treat the exchange and correlation energy
 on the basis of the Adiabatic Connection Fluctuation-Dissipation 
(ACFD) theorem~\cite{Langreth, Langreth2} have raised considerable 
interest in recent years~\cite{PhysRevB.65.235109, Fuchs, Jiang,Hellgren3, Hellgren1,
Nguyen, range-separated, PhysRevLett.103.056401, PhysRevB.81.115126,
jcp.132.044109, Hellgren4, Hebelmann, SE, review-RPA, Thygesen} 
mainly because they provide a route to overcome the shortcomings 
of standard local-density-approximation/generalized-gradient-approximation 
density-functional theory (LDA/GGA DFT). In particular i) an exact 
expression for the exchange-correlation (xc) energy in term of 
density-density response function can be derived from the ACFD 
theorem providing a promising way to develop a systematic improvement 
for the xc functional; ii) all ACFD methods treat the exchange 
energy exactly thus canceling out the spurious self-interaction 
error present in Hartree energy; moreover iii) the correlation 
energy is fully non-local and automatically includes van der Waals 
interactions.

The ACFD method is computationally very demanding and most often it 
is limited to a post self-consistent correction where the xc energy 
is computed from the charge density obtained from a self-consistent 
calculation performed with a more traditional xc functional. The basic 
ingredients needed to compute the correlation energy within the ACFD 
formalism are the density-density response function of the non-interacting 
KS system and the density-density response function of a system where 
the  electron-electron interaction is scaled by a coupling constant. 
While for the former an explicit expression exists, the latter is usually 
calculated from the Dyson equation of time-dependent 
density functional theory\cite{LRTDDFT} containing the xc kernel, $f_{\rm xc}$, 
that needs to be approximated.

%While the former is easily accessible the latter has to be determined approximately via time-dependen-density-functional theory (TDDFT) that is via an approximation of the exchange and correlation kernel appearing in the Dyson equation for the interacting response function.

The random phase approximation (RPA) is the simplest approximation; 
the xc kernel is simply neglected and only the frequency-independent 
Coulomb or Hartree kernel is taken into account. While correctly 
describing van der Waals interactions~\cite{Dobson, Topics} and 
static correlation~\cite{Fuchs,Hebelmann,caruso_bond_2013}, as seen for instance 
when studying H$_2$ dissociation, RPA is known to overestimate 
the correlation energies and thus to poorly describe total 
energies.~\cite{Hellgren3,Jiang}

In this respect various approaches have been developed in order to 
correct the RPA~\cite{RPA+,range-separated,SE}. 
A systematic possibility to address the shortcomings of RPA is to 
include all terms up to a given power of the interaction strength 
in the kernel. To linear order this implies including not only the 
Coulomb kernel, defining RPA, but also an exchange contribution.
%%%%%%%%%%%%%%%%%%%%%%%%%%%%%%%%%%%%%%%%%%%%%%%%%%%%%%%%%%%%%%%%%%%%%%%%%%%%%%%%%%%%%%%%
The frequency-dependent exact-exchange kernel, $f_{\rm x}$, has been derived
by G\"orling~\cite{Gorling2,Gorling1,kim-gorling} from the time-dependent optimized effective potential (TDOEP) method
and by Hellgren and von Barth~\cite{Hellgren1,Hellgren2,Hellgren3} from a variational formulation of many-body
perturbation theory (MBPT). The corresponding 
approximation for the density-density response function, named RPAx, is
obtained by solving the Dyson equation setting $f_{\rm xc} = f_{\rm x}$ and has been successfully 
used in the ACFD formula to compute correlation energies of  atoms~\cite{Hellgren1,Hellgren4}
and molecules.~\cite{hesselman_random_2010,Hebelmann,bleiziffer_RI_2012}

Here we set the RPAx within the context of a general scheme which 
allows to formally define a power expansion of the xc 
kernel combining the general ACFD theory with a many-body approach, 
specifically the G\"orling-Levy Perturbation Theory~\cite{Gorling0} (GLPT), 
along the adiabatic-connection path.
To first order this reduces to the RPAx for which a novel and efficient 
implementation based on an eigenvalue decomposition of the interacting 
time-dependent density response function in the limit of vanishing 
electron-electron interaction, is proposed. 

The performance of the RPAx has in this work been tested on 
the homogeneous electron gas (HEG) at different values of $r_s$ as well 
as on the dissociation of diatomic molecules with covalent bonds such as H$_2$ and N$_2$. 
The results give further support to the accuracy of the RPAx but also reveal
an instability or pathological behavior in the low density regime of the HEG and N$_2$, 
which leads to a break-down of the approximation. This breakdown points to the need for
including correlation or a screening of the exchange kernel. However, we here show
that such a procedure is not always necessary and, in particular, if the aim is to 
calculate total energies. Instead we reduce the effect of the ``bare'' particle-hole interaction by omitting all 
higher order particle-hole scatterings. This can be achieved by expanding the RPA 
response function in the irreducible polarizability, approximated to first order. 
In this way we are able to fix the instability and at the same time keep the overall 
accuracy of the RPAx.

\section{Systematic improvement of the correlation energy}\label{sect:sect1}

Within the ACFD framework a formally exact expression for the 
exchange-correlation energy $E_{\rm xc}$ of an electronic system can 
be derived:~\cite{Langreth,Langreth2}
\begin{align}
  E_{\rm xc}= & -\frac{1}{2} \int_0^1 d\lambda \int d\mathbf{r} d\mathbf{r}' \frac{e^2}{|\mathbf{r}-\mathbf{r}'|} 
 \nonumber \\
 & \times \left\{ \frac{\hbar}{\pi}\int_0^{\infty} \chi_{\lambda}(\mathbf{r},\mathbf{r}' ; iu) \; du + \delta(\mathbf{r}-\mathbf{r}')
   n(\mathbf{r})  \right\}
 \label{eq1.1}
\end{align}
where $\chi_{\lambda}(\mathbf{r},\mathbf{r}' | iu)$ is the density response 
function at imaginary frequency, $iu$, of a system whose electrons interact 
with a scaled Coulomb interaction, $\lambda e^2/|\mathbf{r}-\mathbf{r}'|$, 
and move in a local potential chosen in such a way to keep the electronic 
density fixed to the ground state density of the fully interacting system 
($\lambda =1$).
At $\lambda = 1$, the local potential is equal to the external potential 
(usually the nuclear potential) of the fully interacting system and 
$H_{\lambda=1}$ coincides with the fully interacting Hamiltonian while 
at $\lambda = 0$ the local potential coincides with the KS potential 
and $H_{\lambda=0}$ is the KS Hamiltonian. For intermediate values 
of $\lambda$ the Hamiltonian of the system is~\cite{Gorling0}
\begin{equation} 
 H_{\lambda} = H_{KS} + \lambda  \left( W - \upsilon_{\rm H} - \upsilon_{\rm x} \right) - \frac{\delta E_{\rm c}^{\lambda}}{\delta n}
\label{eq1.2}
\end{equation}
where $\upsilon_{\rm H}$ is the Hartree potential, $\upsilon_{\rm x}$ is 
the local exchange potential and $\delta E_{\rm c}^{\lambda}/\delta n$ is 
the correlation contribution to the potential. The exchange potential is 
defined as the functional derivative of the exact KS exchange energy
\begin{equation}
 E_{\rm x} = -\frac{e^2}{2}\int d\mathbf{r} d\mathbf{r}' \frac{|\sum_i^{occ} \phi_i^*(\mathbf{r}) \phi_i(\mathbf{r}')|^2}
 {|\mathbf{r}-\mathbf{r}'|},
 \label{eq1.2bis}
\end{equation}
that has the same expression as the Hartree-Fock exchange energy but is 
evaluated with the KS orbitals $\phi_i(\mathbf{r})$. It is easy to verify that
it can be derived from Eq.~(\ref{eq1.1}) 
replacing $\chi_{\lambda}$ with the non-interacting density response 
function
\begin{equation}
\chi_0(\mathbf{r},\mathbf{r}';iu) = \sum_{ij}(f_i-f_j)
\frac{\phi_i^*(\mathbf{r}) \phi_j(\mathbf{r}) \phi_j^*(\mathbf{r}') \phi_i\mathbf{r}')}
     {\epsilon_i-\epsilon_j + i\hbar u} 
 \label{eq1.2ter}
\end{equation}
where $\epsilon_i$, $\phi_i(\mathbf{r})$ and $f_i$ are the KS eigenvalues, KS orbitals
and occupation numbers, respectively.  
Subtracting the KS exchange energy from Eq.~(\ref{eq1.1}) 
the correlation energy $E_{\rm c}$ is obtained in terms of linear density 
responses:
\begin{equation}
 E_{\rm c} = -\frac{\hbar}{2\pi} \int_0^1 d\lambda \int_0^{\infty} du\; \text{Tr} \left\{ \upsilon_{\rm c} \left[ \chi_{\lambda}(iu) 
 - \chi_0(iu) \right] \right\} 
  \label{eq1.3}
\end{equation}
where $\upsilon_{\rm c}=e^2/|\mathbf{r}-\mathbf{r}'|$ is the Coulomb kernel 
and $\chi_0(iu)$ is the density response function of the non-interacting KS 
system. For $\lambda > 0$ the interacting density response function 
$\chi_{\lambda}(iu)$ can be related to the non-interacting one 
via a Dyson equation obtained from time-dependent density functional theory (TDDFT):
\begin{equation}
 \chi_{\lambda}(iu) = \chi_0(iu) + \chi_0(iu) [ \lambda \upsilon_{\rm c} + f^{\lambda}_{\rm xc}(iu)] \chi_{\lambda}(iu)
\label{eq1.4}
\end{equation}
where $f^{\lambda}_{\rm xc}(iu)$ is the scaled frequency-dependent xc kernel. 
Spatial coordinates dependence is implicit in the matrix notation.
When the xc kernel is specified one can thus determine a corresponding 
correlation energy via Eq. (\ref{eq1.3}).

%Untill recently quite little was known about the xc-kernel. 
%The most common approximations to $f_{\rm xc}$, are based on properties 
%of the HEG~( see, for instance, Refs.~\onlinecite{Topics},~\onlinecite{Giuliani-Vignale}
 %and~\onlinecite{lein_electron_2000} for a review), the simplest being the Adiabatic Local 
%Density Approximation (ALDA) where non-locality and frequency dependence of
%the kernel are completely neglected. A truly non-local (but still adiabatic) 
%approximation for $f_{\rm xc}$ has been obtained by Petersilka \emph{et al.}\cite{petersilka} 
%starting from the time-dependent generalization of the approximate OEP scheme
%proposed by Krieger, Li, and Iafrate~\cite{KLI}. 

%Recently the expression for the full frequency-dependent exact-exchange 
%contribution to $f_{xc}$ has been derived by G\"orling~\cite{Gorling2,Gorling1,kim-gorling}
%within the TDOEP framework and 
%by Hellgren and von Barth~\cite{Hellgren2,Hellgren1,Hellgren4} from a variational 
%formulation of  MBPT. It defines, together to the Coulomb kernel, the full first order 
%contribution to $f_{\rm xc}$ and, if plugged into the Dyson equation~(\ref{eq1.4}),
%leads to the RPAx approximation for the response function which
%is exact up to first order in the interaction strength $\lambda$.

In the following we will describe a general scheme
which allows us to compute the xc kernel to a given order, 
thus establishing a link between the TDDFT expression for the response 
function in Eq.~(\ref{eq1.4}) and  the power expansion of $\chi_{\lambda}$ 
in the interaction strength, which can be obtained resorting to the well established 
GLPT~\cite{Gorling0} along the adiabatic-connection path.

Considering the power expansion for the xc kernel $f_{\rm xc}^{\lambda} = 
\lambda f_{\rm x} + \lambda^2 f_{\rm c}^{(2)} + \dots$ and explicitly 
expanding the Dyson equation~(\ref{eq1.4}) in power of the interaction 
strength
\begin{align}
 \chi_{\lambda} = & \chi_0 +\lambda \left[ \chi_0 \left( \upsilon_{\rm c} + f_{\rm x} \right) \chi_0 \right] + 
\nonumber \\
+ & \lambda^2 \left[ \chi_0 \left( \upsilon_{\rm c} + f_{\rm x} \right) \chi_0\left( \upsilon_{\rm c} + f_{\rm x} \right) \chi_0
 + \chi_0 f_{\rm c}^{(2)} \chi_0 \right] + \ldots,
\label{eq1.5}
\end{align}
it can be seen that the first order kernel, $\upsilon_{\rm c}+f_{\rm x}$, 
is intimately related to the first order variation of $\chi_{\lambda}$ with 
respect to $\lambda$ and similarly higher order correlation contributions 
to the kernel are related to the corresponding power in the $\chi_{\lambda}$
expansion. 

Therefore i) we can define an arbitrarily accurate approximation to the 
density-density response function considering the expansion of the kernel 
up to a desired order in $\lambda$:
\begin{equation}
 \chi_{\lambda}^{(n)} = \chi_0 + \chi_0 [\lambda\upsilon_{\rm c} + \lambda f_{\rm x} + \ldots +\lambda^n f_{\rm c}^{(n)}]\chi_{\lambda}^{(n)};
\label{eq1.6}
\end{equation}
where ii) the kernel up to order $\lambda^n$ can be exactly determined by 
comparing with the $\lambda^{n}$ expansion of $\chi_{\lambda}$ from GLPT 
and iii) the solution of the Dyson equation for $\chi_{\lambda}^{(n)}$ leads 
to a density-density response function which is exact to order $\lambda^n$ 
but also contains higher-order terms.

In order to solve the Many Body Hamiltonian in Eq.~(\ref{eq1.2}) so as to 
obtain the xc kernel to a given order in $\lambda$, the xc potential, and 
hence the xc energy, must be known up to the same level. This apparent 
circular dependence does not actually hinder the application of the 
procedure since, thanks to the coupling constant integration involved 
in Eq.~(\ref{eq1.3}), the knowledge of the xc energy, and therefore its 
functional derivatives, up to order $\lambda^n$ only depends on the xc 
kernel up to order $\lambda^{n-1}.$ 
Our strategy can thus be applied in the sequential way  
\begin{align}
& E_0 \xrightarrow{\delta/\delta n} \upsilon_{KS} \xrightarrow{GLPT} \chi_0 \xrightarrow{ACFD} 
\nonumber \\
\rightarrow & E_{\rm x} \xrightarrow{\delta/\delta n} \upsilon_{\rm x} \xrightarrow{GLPT} \left( f_{\rm x}, \chi^{(1)}_{\lambda} \right) \xrightarrow{ACFD}
\nonumber \\ 
\rightarrow & E_{\rm c}^{(r2)}  \rightarrow \ldots
%\xrightarrow{\delta/\delta n} \upsilon_{\rm c}^{(2)} \xrightarrow{MBPT} \left( f_{\rm c}^{(2)} , \chi^{(2)}_{\lambda} \right) \xrightarrow{ACFD}
%\nonumber \\ 
%\rightarrow & E_{\rm c}^{(3)} \rightarrow \ldots
\label{eq1.7}
\end{align}
showing that to $0^{\rm th}$ order, i.e. replacing $\chi_{\lambda}$ with its 
non-interacting counterpart $\chi_0$, the exact-exchange KS energy is obtained;
moving to the next step, the exact-exchange kernel can be derived from first order GLPT
and we recover the so-called RPAx approximation for the response function, i.e. $\chi_{\lambda}^{(1)}$,
and for the correlation energy, i.e. $E_{\rm c}^{(r2)}$. 
Notice that the RPAx correlation energy $E_{\rm c}^{(r2)}$ is exact to order 
$\lambda^2$ but also contains, although in an approximate way, 
all higher-order terms, and should not be confused 
with the $2^{\rm nd}$ perturbative correction to the correlation energy 
in the G\"orling-Levy perturbation theory~\cite{Gorling0}.
 
The mathematical complexity of this sequential procedure
increases very rapidly and makes extremely hard its application already 
at the second order; nevertheless the prescription is in principle 
given. The functional derivative of $E_{\rm c}^{(r2)}$ with 
respect to the density defines the exact $\lambda^2$ correction to the 
Hamiltonian in Eq.~(\ref{eq1.2}) and allows to apply the GLPT to 
second order and hence to have access to corresponding second-order contribution
to the xc kernel. Solving the Dyson equation with the improved kernel 
defines a new approximation for the response function $\chi_{\lambda}^{(2)}$
which is exact up to second order.
Plugging $ \chi_{\lambda}^{(2)}$ into
the ACFD formula~(\ref{eq1.3}), leads to a new approximation for the correlation 
energy, $E_{\rm c}^{(3r)}$, which is exact to order $\lambda^3$ but also contains, 
although in an approximate way, all higher-order terms.

Essentially this scheme can be regarded as a revised version of the standard
GLPT~\cite{Gorling0,Gorling0-td,Gorling1} with the additional step provided 
by the solution of the Dyson equation for the response function and the calculation 
of a non-perturbative correlation energy (all order in the coupling-constant appears 
in $E_{\rm c}^{(r2)}$ and following approximation to $E_c$) from the ACFD formula in Eq.~(\ref{eq1.3}). 
In this way we expect this approach to be applicable also to small gap or 
metallic systems where finite-order many-body perturbation theories break 
down~\cite{Gell-Mann,Giuliani-Vignale}.

Having introduce the general framework, we apply our strategy to first order in the coupling 
strength, hence we focus on the frequency-dependent exact-exchange kernel 
$f_{\rm x}$ and on the calculation of the contribution $E_{\rm c}^{(r2)}$ to the correlation energy
(previously denoted as RPAx~\cite{Hellgren1,Hellgren4} or EXXRPA~\cite{bleiziffer_RI_2012,Hebelmann})
for which we propose a novel and efficient implementation.

\section{Efficient calculation of RPAx correlation energy}
Our implementation for computing the RPAx correlation energy is based on an eigenvalue 
decomposition of the time-dependent response function $\chi_{\lambda}$ in the limit of 
vanishing coupling constant. The scheme described below is a generalization of the 
implementation proposed by Nguyen and de Gironcoli~\cite{Nguyen} for computing RPA 
correlation energies.

\subsection{RPAx Correlation Energy}\label{sect:sect2}
Let us start by defining the following generalized eigenvalue problem:
\begin{equation}
-\chi_0[\upsilon_{\rm c}+f_{\rm x}]\chi_0 |\omega_{\alpha}\rangle = a_{\alpha} [-\chi_0] |\omega_{\alpha}\rangle 
 %\quad  \langle \omega_{\alpha} | [-\chi_0] | \omega_{\beta}\rangle = \delta_{\alpha\beta}
 \label{eq2.1.1}
\end{equation}
where the eigenpairs $\{|\omega_{\alpha}\rangle,a_{\alpha}\}$ and all the operators 
implicitly depend on the imaginary frequency $iu$. Once the solution of the generalized 
eigenvalue problem (\ref{eq2.1.1}) is available, 
%the RPAx Dyson-like equation is solved
%\begin{align}
%& \chi_{\lambda} | \omega_{\alpha} \rangle = 
%\chi_0 | \omega_{\alpha} \rangle + \lambda \chi_{\lambda} [\upsilon_{\rm c}+f_{\rm x}] \chi_0 | \omega_{\alpha} \rangle 
%\quad \Rightarrow \quad \nonumber \\
%& \quad \Rightarrow \quad \chi_{\lambda} | \omega_{\alpha} \rangle = \frac{\chi_0 | \omega_{\alpha} \rangle}{1-\lambda a_{\alpha}},
%\label{eq2.1.2}
%\end{align}
%(ii) the traces in Eq.~(\ref{eq1.3}) are simply computed as the sum over the diagonal elements on the basis defined by the eigenpotentials $\{|\omega_{\alpha}\rangle\}$ 
the trace in Eq.~(\ref{eq1.3}) is simply given by 
\begin{align}
 &\text{Tr}\left[ \upsilon_{\rm c} \left( \chi_{\lambda}-\chi_0\right) \right] = 
% \nonumber \\
  \sum_{\alpha} \left( 1- \frac{1}{1-\lambda a_{\alpha}}\right) \langle \omega_{\alpha} | \chi_0 \upsilon_{\rm c} \chi_0 | \omega_{\alpha} \rangle
 \label{eq2.1.2}
\end{align}
and the integration over the coupling constant can be calculated analytically, 
leading to the final expression
\begin{align}
 E_{\rm c}^{(r2)} =  -\frac{\hbar}{2\pi} & \int_0^{\infty} du\; 
 \sum_{\alpha} \frac{\langle \omega_{\alpha}| \chi_0\upsilon_{\rm c}\chi_0 | \omega_{\alpha} \rangle}{a_{\alpha}(iu)} 
 \nonumber \\
 & \times \left\{ a_{\alpha}(iu) + \ln[1-a_{\alpha}(iu)]\right\}
 \label{eq2.1.3}
\end{align}
Notice that Eq.~(\ref{eq2.1.1}) and (\ref{eq2.1.3}) demonstrate that knowledge 
of $\chi_0 f_{\rm x} \chi_0$ is sufficient for computing the RPAx correlation energy and the 
exact-exchange kernel alone is not needed.

\subsection{Exact-Exchange kernel}\label{sect:sect2.1}

The exact expression for $h_{\rm x} = \chi_0f_{\rm x}\chi_0$ in term of the KS eigenvalues and 
eigenfunctions has been derived by G\"orling starting from the time-dependent 
optimized potential method equation\cite{Gorling2} and by Hellgren and 
von Barth starting from the variational formulation of many-body perturbation 
theory\cite{Hellgren1,Hellgren3}. 
Here we propose an alternative derivation staying within the general scheme 
described in the previous section.

In section \ref{sect:sect1} it has 
been shown that $h_{\upsilon \rm{x}} = \chi_0(\upsilon_{\rm c} +f_{\rm x})\chi_0$ is the first 
order correction to the non-interacting response function $\chi_0$ due to the 
switching on of the perturbation 
%$\delta \hat{V}^{(1)} = \hat{W} + \frac{d \hat{\upsilon}_{\lambda}}{d\lambda}|_{_{\lambda=0}} \equiv \hat{W} -\hat{\upsilon}_H - \hat{\upsilon}_x$
 $\delta \hat{V} = \hat{W} -\hat{\upsilon}_H - \hat{\upsilon}_x$.
%(only linear term in $\lambda$ is needed). 
Moreover in the previous sub-section it has been shown that the eigenvalues and 
eigenvectors of $h_{\upsilon \rm{x}}$ are sufficient for computing RPAx correlation 
energies. In what follows we derive the exact expression for the matrix elements 
of $h_{\upsilon \rm{x}}$ in term of the KS eigenvalues and eigenfunctions and their 
first order corrections only, and show how they can be efficiently computed 
resorting to the linear-response techniques of density functional perturbation 
theory~\cite{DFPT}.

Let us start by considering the matrix element of $\chi_0$ on two arbitrary, $\alpha$ and $\beta$, time-dependent perturbing potentials
$\Delta V = \Delta V(\mathbf{r}) e^{ut}$ at imaginary frequency $\omega = iu$
\begin{align} 
\chi_0^{\alpha\beta}(iu) &= \langle \Delta^{\alpha} V | \chi_0 | \Delta^{\beta} V \rangle \nonumber \\ 
& = \int d^3 \mathbf{r} \, \Delta^{\alpha} V(\mathbf{r}) \Delta^{\beta} n(\mathbf{r};iu)
\label{eq2.5}
\end{align}
For a non degenerate ground state the linear response density $\Delta n$ at 
imaginary frequency $\omega = iu$ can be written as
\begin{equation}
  \Delta n(\mathbf{r};iu) =\langle \Phi_0 | \hat{n}(\mathbf{r}) | \Delta \Phi_0^{(+)} + \Delta \Phi_0^{(-)} \rangle
\label{eq2.6}
\end{equation}
 where $|\Delta \Phi_0^{\pm}\rangle$ are the first order corrections to the 
KS wavefunction $|\Phi_0\rangle$ due to the perturbation 
$\Delta V$ and satisfy the linearized time-dependent KS equations
\begin{equation}
 [H_{KS}-(E_0 \pm iu)]  | \Delta \Phi_0^{(\pm)} \rangle + \Delta V | \Phi_0 \rangle = 0.
 \label{eq2.7}
\end{equation}
Eq.~(\ref{eq2.5}) becomes $\chi_0^{\alpha\beta}(iu) = \langle \Phi_0 | \Delta^{\alpha} V | \Delta^{\beta} \Phi_0^{(+)} + \Delta^{\beta} \Phi_0^{(-)} \rangle
$
and if the (static) perturbation $\delta V$ is turned on, the first order 
correction to $\chi_0$, i.e. $h_{\upsilon \rm{x}}$, in the coupling constant 
$\lambda$ can be computed:
%just varying Eq.~(\ref{eq2.8})
\begin{align}
 h_{\upsilon \rm{x}}^{\alpha\beta} = \delta\chi_0^{\alpha\beta} & = \langle \delta \Phi_0 |\Delta^{\alpha} V |
\Delta^{\beta} \Phi_0^{(+)} + \Delta^{\beta} \Phi_0^{(-)} \rangle \nonumber \\ 
 & + \langle \Phi_0 | \Delta^{\alpha} V | \delta \Delta^{\beta} \Phi_0^{(+)} + \delta \Delta^{\beta} \Phi_0^{(-)} \rangle
\label{eq2.9}
 \end{align}  
where $|\delta \Delta \Phi_0 \rangle$ is obtained by taking the linear variation of Eq~(\ref{eq2.7})
\begin{align}
 & \left[ H_{KS}-(E_0 \pm iu) \right]|\delta\Delta \Phi_0^{(\pm)}\rangle + \nonumber \\
+ & \left[\delta V -\delta E_0\right]|\Delta \Phi_0^{(\pm)}\rangle+\Delta V|\delta \Phi_0\rangle=0
 \label{eq2.10}
\end{align}
while the static correction vector $|\delta \Phi_0 \rangle$ satisfies the linearized 
time-independent Schr\"odinger equation
\begin{equation}
 [H_{KS}-E_0]  | \delta \Phi_0 \rangle + [\delta V -\delta E_0]| \Phi_0 \rangle = 0
 \label{eq2.9bis}
\end{equation}
with $\delta E_0 = \langle \Phi_0 | \delta V | \Phi_0 \rangle$.

With a simple manipulation it's easy to show that $\delta \chi_0$  depends only on the 
GS wavefunction and its first order corrections (and not on the second order correction 
$|\delta \Delta \Phi_0 \rangle$). Taking the h.c. of Eq.~(\ref{eq2.7}) and multiplying 
it on the right  by $|\delta\Delta\Phi_0 \rangle$ and Eq.~(\ref{eq2.10}) on the left by
$\langle \Delta \Phi_0 |$ and subtracting the two identities so obtained, an expression 
for $\langle \Phi_0 | \Delta^{\alpha} V | \delta \Delta^{\beta} \Phi_0^{(-)} + \delta \Delta^{\beta} \Phi_0^{(+)} \rangle$ is obtained where the second order corrections cancel out.

The final expression for $h_{\upsilon \rm{x}}^{\alpha \beta}$ becomes:
\begin{align}
  h_{\upsilon \rm{x}}^{\alpha \beta} &= \langle \Delta^{\alpha} \Phi_0^{(-)}+  \Delta^{\alpha} \Phi_0^{(+)} | \Delta^{\beta}
V | \delta \Phi_0 \rangle \nonumber \\  
 &+ \langle \delta \Phi_0 |\Delta^{\alpha} V | \Delta^{\beta} \Phi_0^{(+)} + \Delta^{\beta} \Phi_0^{(-)} \rangle  \cr
 &  - \left[ \langle \Delta^{\alpha} \Phi_0^{(+)} | \Delta^{\beta} \Phi_0^{(-)} \rangle  
      +  \langle \Delta^{\alpha} \Phi_0^{(-)} | \Delta^{\beta} \Phi_0^{(+)} \rangle \right] 
%        \langle \Phi_0 | \delta V | \Phi_0 \rangle  
         \delta E_0 \cr
 & +  \langle \Delta^{\alpha} \Phi_0^{(+)} | \delta V | \Delta^{\beta} \Phi_0^{(-)} \rangle +
        \langle \Delta^{\alpha} \Phi_0^{(-)} | \delta V | \Delta^{\beta} \Phi_0^{(+)} \rangle. \cr
\label{eq2.12}
\end{align}
Eq.~(\ref{eq2.12}) together with Eq.~(\ref{eq2.7}) and Eq.~(\ref{eq2.9bis}) defines 
the matrix elements $h_{\upsilon \rm{x}}^{\alpha \beta}$ as a function of the KS many-body 
ground-state wavefunctions $|\Phi_0\rangle$ and its first order corrections 
$|\Delta \Phi_0^{\pm}\rangle$ and $|\delta \Phi_0\rangle$.
Introducing their definitions in terms of the single particle KS orbitals, 
$\phi_{a}$'s, and their first order variations, $\Delta \phi^{(\pm)}_a$'s and 
$\delta\phi_a$'s,  Eq.~(\ref{eq2.12}) becomes:
\begin{widetext}
\begin{align}
% \langle \Delta^{\alpha} V & | \chi_0 [\upsilon_{\rm c}+f_{\rm x}] \chi_0 | \Delta^{\beta} V \rangle =
  h_{\upsilon \rm{x}}^{\alpha\beta} =
&
 +\sum_{ab} \langle \Delta^{\alpha} \phi_a^{(-)} \phi_b | W | \Delta^{\beta} \phi_b^{(+)} \phi_a \rangle
 +\sum_{ab} \langle \Delta^{\alpha} \phi_a^{(+)} \phi_b | W | \Delta^{\beta} \phi_b^{(-)} \phi_a \rangle 
 \nonumber \\
 & 
 +\sum_{ab} \langle \Delta^{\alpha} \phi_a^{(-)} \phi_b  | W | \Delta^{\beta} \phi_b^{(-)} \phi_a \rangle
 +\sum_{ab} \langle \Delta^{\alpha} \phi_a^{(+)} \phi_b  | W | \Delta^{\beta} \phi_b^{(+)} \phi_a \rangle
\nonumber \\
&
 -\sum_{ab} \langle \Delta^{\alpha} \phi_a^{(-)} \phi_b | W | \phi_a \Delta^{\beta} \phi_b^{(+)} \rangle
 -\sum_{ab} \langle \Delta^{\alpha} \phi_a^{(+)} \phi_b | W | \phi_a \Delta^{\beta} \phi_b^{(-)} \rangle 
 \nonumber \\
 & 
 -\sum_{ab} \langle \phi_b \phi_a  | W | \Delta^{\beta} \phi_b^{(+)} \Delta^{\alpha} \phi_a^{(-)} \rangle
 -\sum_{ab} \langle \phi_b \phi_a  | W | \Delta^{\beta} \phi_b^{(-)} \Delta^{\alpha} \phi_a^{(+)} \rangle
\nonumber \\
&
 + \sum_{a} \langle \Delta^{\alpha} \phi_a^{(-)} | V_x-v_x | \Delta^{\beta} \phi_a^{(+)} \rangle 
 + \sum_{a} \langle \Delta^{\alpha} \phi_a^{(+)} | V_x-v_x | \Delta^{\beta} \phi_a^{(-)} \rangle 
\nonumber \\
&
 - \sum_{ab} \left[ {\langle \Delta^{\alpha} \phi_a^{(-)} | \Delta^{\beta} \phi_b^{(+)} \rangle + 
           \langle \Delta^{\alpha} \phi_a^{(+)} | \Delta^{\beta} \phi_b^{(-)} \rangle } \right]
 \langle \phi_b | V_x-v_x | \phi_a \rangle 
\nonumber \\
&
 +\sum_a \langle \delta \phi_a | \Delta^{\alpha} V^* | \Delta^{\beta} \phi_a^{(+)} + \Delta^{\beta} \phi_a^{(-)} \rangle 
 +\sum_a \langle \Delta^{\alpha} \phi_a^{(+)} + \Delta^{\alpha} \phi_a^{(-)} | \Delta^{\beta} V | \delta \phi_a \rangle
\nonumber \\
&
-\sum_{ab} \langle \delta \phi_a | \Delta^{\beta} \phi_b^{(+)} + \Delta^{\beta} \phi_b^{(-)} \rangle \langle \phi_b|\Delta^{\alpha} V^* | \phi_a \rangle
- \sum_{ab} \langle \Delta^{\alpha} \phi_a^{(-)} + \Delta^{\alpha} \phi_a^{(+)} | \delta \phi_b \rangle  \langle \phi_b|\Delta^{\beta} V | \phi_a \rangle
\label{eq2.13}
\end{align}
\end{widetext}
where the sums run over the occupied single-particle KS state only and 
$|\Delta\phi^{(\pm)}_a\rangle$ and $|\delta\phi_a\rangle$ are the 
(conduction-band projected) variations of the occupied single-particle state. 
They can be efficiently computed resorting to the linear-response techniques 
of density functional perturbation theory~\cite{DFPT}:
\begin{align}
 & [H^0+ \gamma P_{\upsilon} -(\varepsilon_a \pm iu) ]|\Delta\phi_a^{\pm}\rangle = -(1-P_{\upsilon})\Delta V |\phi_a\rangle
 \nonumber \\
 & [H^0+ \gamma P_{\upsilon} - \varepsilon_a]|\delta\phi_a\rangle = -(1-P_{\upsilon})[V_x-\upsilon_{\rm x}] |\phi_a\rangle
 \label{eq2.14}
\end{align}
where $V_x$ is the non-local exchange operator identical to the Hartree-Fock 
one but constructed from KS orbitals, $P_{\upsilon}=\sum_a^{occ}|\phi_a\rangle\langle\phi_a|$ 
is the projector on the occupied manyfold and $\gamma$ is a positive constant 
larger than the valence bandwidth in order to ensure that the linear system is 
not singular even in the limit for $iu \rightarrow 0$.

Inserting the formal solutions for $|\Delta\phi_a^{(\pm)}\rangle$ and $|\delta\phi_a\rangle$
from Eq.~(\ref{eq2.14}) into Eq.~(\ref{eq2.13}), removing the trial perturbing potential $\Delta V$ and
sending $iu \rightarrow \omega$, the expression for $h_x(\mathbf{r},\mathbf{r}'; \omega)$
previously derived in Ref.~\onlinecite{Gorling2,bleiziffer_RI_2012} and in Ref.~\onlinecite{Hellgren2} is recovered.

The scheme described above has been implemented in the \textit{Quantum ESPRESSO} 
distribution~\cite{QE}. The basic operations involved in the calculation 
of the matrix elements $h_{\upsilon \rm{x}}^{\alpha\beta}$ are the same required 
for the calculation of RPA energy and potential in the implementations proposed 
by Nguyen and de Gironcoli~\cite{Nguyen} and Nguyen \emph{et al.}~\cite{Linh} 
respectively, meaning that our RPAx calculation has a computational cost 
comparable to their RPA implementations and maintains their favorable scaling.

\section{Homogeneous electron gas}
As a test of the accuracy of the new approximation we choose the simple 
homogeneous electron gas. The homogeneous electron gas is an idealized system 
of electrons moving in a uniform neutralizing background. At zero temperature 
it is characterized by two parameters only, i.e. the number density 
$n=1/(4\pi r_s^3 a_B^3/3)$, or equivalently the Wigner-Seitz radius $r_s$,
and the spin polarization $\zeta=|n^{\uparrow} - n^{\downarrow}| / (n^{\uparrow} + n^{\downarrow})$,
where $n^{\uparrow(\downarrow)}$ is the density of spin up (down) electrons 
and $n=n^{\uparrow} + n^{\downarrow}$. Despite its simplicity, (i) the HEG 
model represents the first approximation to metals where the valence electrons 
are weakly bound to the ionic cores, (ii) the system is found to display a 
complex phase diagram including transition to the Wigner crystal and in 
addition (iii) it provides the basic ingredient of any practical density 
functional calculation. The most widely used approximations for the unknown 
xc-energy functional are based on properties of the HEG.

\subsection{Unpolarized HEG}\label{sect:sect3.1}
We begin by studying the unpolarized HEG.
While the solution of Dyson equation is demanding in general, it becomes 
trivial in the case of the HEG; the response functions and the kernels are 
all diagonal in momentum space and the RPAx Dyson equation can be easily solved as
\begin{equation}
 \chi_{\lambda}(q,iu) = \frac{\chi_0(q,iu)}{1-\lambda[\upsilon_{\rm c}(q)+f_{\rm x}(q,iu)]\chi_0(q,iu)}
\label{eq3.1}
\end{equation}
where $\upsilon_{\rm c}(q)=4\pi e^2/q^2$ and $f_{\rm x}(q,iu)$ is the exchange kernel 
at a given momentum and frequency.
 
\begin{figure}[t]
\begin{center}
\includegraphics[scale=0.3]{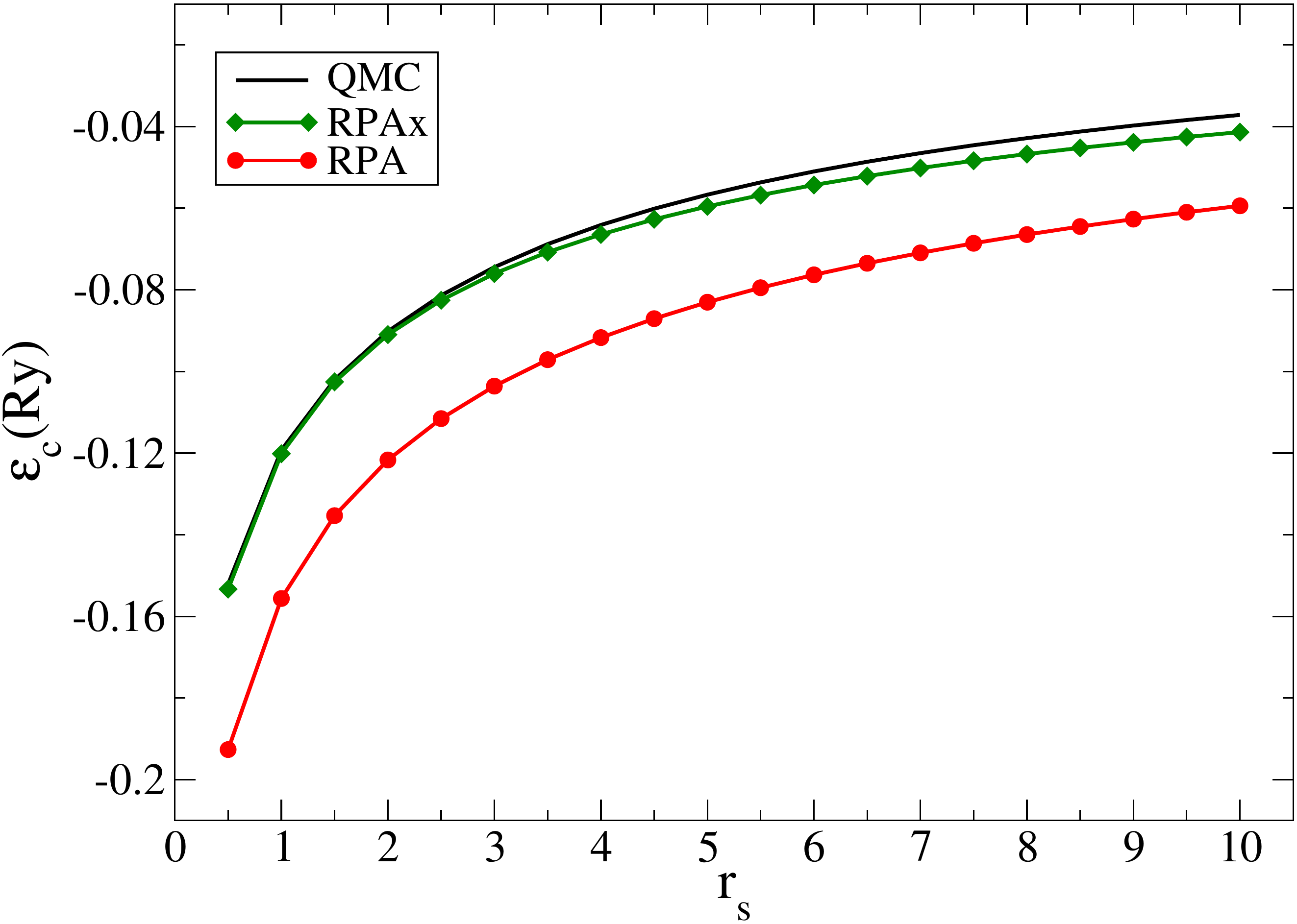}
\caption{Correlation energy per particle in the homogeneous electron gas as
 a function of the Wigner-Seitz radius evaluated with different kernels: 
 RPA (red squares), RPAx (green diamonds) and QMC calculation (black circles). }
\label{fig1}
\end{center}
\end{figure}

The correlation energy per electron $\epsilon_c$ follows from Eq.~(\ref{eq1.3})
 where the trace has been replaced by an integral over momentum $q$ and the 
integration over $\lambda$ has been done analytically
\begin{align}
 \epsilon_c = \frac{\hbar}{2\pi^2 n}  \int_0^{\infty} q^2 dq  & \int_0^{\infty} du \; 
\upsilon_{\rm c}(q)\chi_0(q,iu)  \nonumber \\
 & \times \left[1+ \frac{\ln [1-K(q,iu)]}{K(q,iu)}  \right].
\label{eq3.2}
\end{align}
\begin{nobreak}
Here $K(q,iu)$ has been defined as
\begin{align}
 K(q,iu) &=  [\upsilon_{\rm c}(q) + f_{\rm x}(q,iu)] \chi_0(q,iu) \nonumber  \\
         &= \upsilon(q)\chi_0(q,iu) +\frac{h_x(q,iu)}{\chi_0(q,iu)}
\label{eq3.3}
\end{align}
\end{nobreak}
While the Lindhard function $\chi_0(q,iu)$ at imaginary frequency $iu$ is 
known exactly~\cite{Giuliani-Vignale}, the function $h_x(q,iu)$ can be 
directly derived from the general expression given in Eq.~(\ref{eq2.13}) 
and is given by a six-fold integral over crystal momenta. Its static 
values was computed first numerically by several author 
\cite{Geldart, Antoniewicz, Chevary} and later analytically by Engel 
and Vosko~\cite{EV}.The frequency dependence of $h_x$ has been calculated 
by Bronsens, Lemmens and Devreese~\cite{Bronsens1, Bronsens2} for real 
frequencies and by Richardson and Ashcroft~\cite{Richardson} for imaginary 
frequencies. Following Bronsens \textit{et al.} four integrations can be done 
analytically using cylindrical coordinates; we used numerical quadrature 
for the two remaining integrations. Our numerical integration is able to 
recover the analytic results of Engel and Vosko~\cite{EV} in the limit $u\rightarrow0$.
Finally the integration over momentum $q$ and imaginary frequency $u$ 
in Eq.~(\ref{eq3.2}) has been computed numerically. The results are listed 
in Table~\ref{tab1} and Fig.~\ref{fig1}. RPA can be easily obtained from 
Eq.~(\ref{eq3.2}) and Eq.~(\ref{eq3.3}) with $h_x=0$ and can be seen to 
seriously overestimate the correlation energy at all densities. Including 
the exact exchange kernel greatly improves over simple RPA and the RPAx 
correlation energy per particle is close to the accurate Quantum Monte 
Carlo (QMC) results~\cite{Ceperly}.

\begin{figure}[b]
\begin{center}
\includegraphics[scale=0.3]{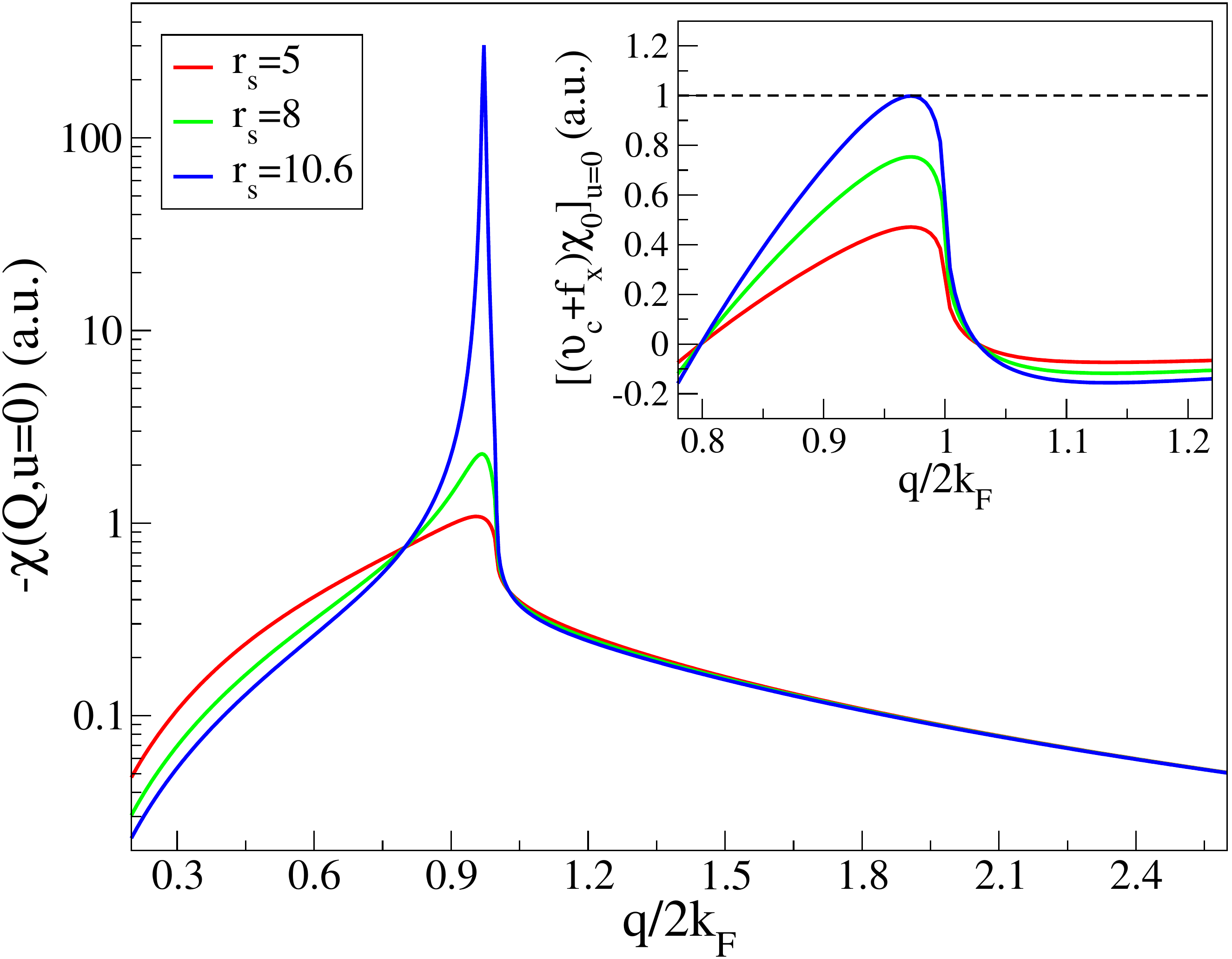}
\caption{Critical behavior of the static density-density RPAx response 
function when the density decreases. For $r_s>10.6$ the system becomes 
unstable respect to charge modulation with wavevector $\approx 2k_F$.}
\label{fig2}
\end{center}
\end{figure}

As expected RPAx works well for small values of $r_s$ and becomes less 
accurate when $r_s$ increases. According to our calculation, within RPAx 
for $r_s > 10.6$ there is a charge density instability with 
wave-vector $q \approx 2k_F$. 
In Fig.~\ref{fig2} the critical behavior of the static density-density RPAx 
response function is shown for the full interacting system (Eq.~(\ref{eq3.1}), 
$\lambda=1$). When the density decreases a pronounced peak appears at $q \approx 2k_F$ 
indicating the instability with respect to charge modulations with this 
wave-vector. As can be seen from the inset in Fig.~(\ref{fig2}), for 
sufficient large values or $r_s$, $K=(\upsilon_{\rm c}+f_{\rm x})\chi_0$ approaches 
unity and the denominator in Eq.~(\ref{eq3.1}) tends to vanish leading 
to the appearance of the peak. 
Beyond $r_s = 10.6$, $K$ exceeds unity and RPAx approximation breaks 
down as the density-density response function $\chi_{\lambda}$ is not 
anymore negative definite.

This instability resembles the charge density wave instability, 
already observed at the Hartree-Fock level by 
Overhauser~\cite{Overhauser,Giuliani-Vignale} and is an artifact of the 
truncation of the kernel expansion to first order in the interacting 
strength. A full treatment of correlation in the QMC calculations 
moves the density instability toward the Wigner crystal to much smaller 
densities corresponding to $r_s \approx 80$.~\cite{Ceperly}

\begin{center}
\begin{table}[t]
\begin{tabular}{c c c c c}
\hline
\hline
\rule[-4mm]{0mm}{1.cm}
 \makebox[1.5cm][c]{$r_s$}   & \makebox[1.5cm][c]{RPA}   &  \makebox[1.5cm][c]{RPAx}  & \makebox[1.5cm][c]{QMC}  \\
\hline
\rule{0mm}{0.4cm}
 0.5   &   -0.194  &  -0.154  &  -0.153  \\
 1.0   &   -0.157  &  -0.121  &  -0.119  \\
 3.0   &   -0.105  &  -0.077  &  -0.074  \\
 5.0   &   -0.085  &  -0.060  &  -0.056  \\
 8.0   &   -0.068  &  -0.047  &  -0.043  \\
%\rule[-0.2cm]{0mm}{0.2cm}
 10.0  &   -0.061  &  -0.042  &  -0.037  \\
%\rule[-0.2cm]{0mm}{0.2cm}
 11.0  &   -0.058  &    --    &  -0.035  \\
\hline
\hline
\end{tabular}
\caption{Corelation Energy per particle with different kernels: 
 RPA ($f^{\lambda}_{xc}=0$), RPAx ($f^{\lambda}_{xc}=\lambda f_{\rm x}$), 
%RPAx plus Adiabatic Local Density Correlation ($f^{\lambda}_{xc}=\lambda f_{\rm x} + f_c^{\lambda,ALDA}$)
 and Quantum Monte Carlo calculation~\cite{Ceperly}. All energies are in Rydberg.}
\label{tab1}
\end{table}
\end{center}

\subsection{Alternative RPAx resummations}\label {sect:sect-alt}
In Sect.~\ref{sect:sect1} we have established a strategy for a systematic 
improvement of the xc kernel. However, because of the complexity of the procedure,
rather than proceeding along this way we propose here
two simple modifications to the original RPAx approximation
which are able to fix the instability problem and, at the same time, to 
give correlation energies on the same level of accuracy as RPAx. 

%We notice that i) the RPA response function is negatively defined for any 
%values of $r_s$ and also that  ii) the RPAx response function is exact 
%up to first order in the coupling strength $\lambda$, meaning that 
%the instability observed at the RPAx level must be due to the re-summation 
%up to infinite order of the ``bare'' exact-exchange kernel contributions, 
%that is without any correlation corrections. Therefore, in order to mitigate 
%the instability problem, we can either improve the kernel by including higher order correlation 
%contributions (rigorous but expensive) or define different re-summations of the 
%exact-exchange kernel (approximate but inexpensive) mimicking the effect of the 
%missing correlation kernel as discussed below.

Introducing the irreducible polarizability $P_{\lambda}$, it is possible to write the interacting 
response function $\chi_{\lambda}$ as\cite{Giuliani-Vignale} 
\begin{equation}
\chi_{\lambda} = P_{\lambda} +\lambda P_{\lambda} \upsilon_{\rm c} \chi_{\lambda}.
\end{equation}
where  
$P_{\lambda} = \chi_0 + \chi_0[\lambda f_{\rm x} + f_c(\lambda)]P_{\lambda}$. Neglecting 
$f_c(\lambda)$ and summing up to infinite order leads again to the RPAx approximation defined above. 
If we instead replace $P_{\lambda}$ with only its first order expansion we can define a new 
approximation, here named tRPAx, which contains only a subset of the original 
RPAx expansion:
\begin{equation}
 \chi^{\rm tRPAx}_{\lambda} =  P_{\lambda}^{(1)} + \lambda P_{\lambda}^{(1)} \upsilon_{\rm c} \chi^{\rm tRPAx}_{\lambda}
 \label{eq3.4}
\end{equation}
with $ P_{\lambda}^{(1)} = \chi_0 + \lambda \chi_0 f_{\rm x} \chi_0 = \chi_0 +\lambda  h_{\rm x}$. In this way
we are only including terms which contain first order particle-hole interactions. 
\begin{figure}[t]
\begin{center}
\includegraphics[scale=0.3]{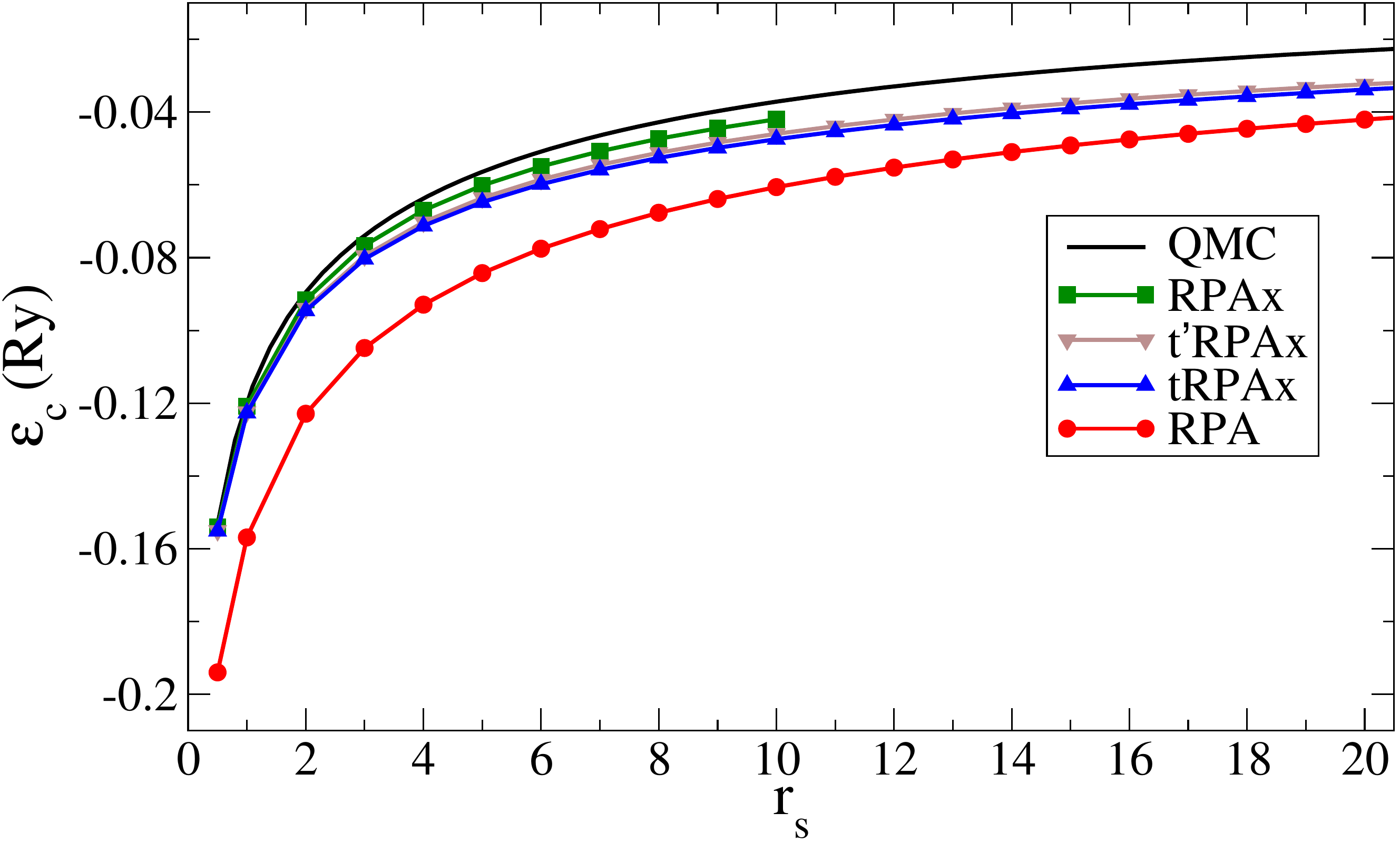}
\caption{Correlation energies per particle as a function of $r_s$ evaluated from the RPA, original 
         and modified RPAx response functions and compared to accurate QMC calculations. }
\label{fig2.2}
\end{center}
\end{figure} 

A similar idea has been proposed in Ref.~\onlinecite{RPAx-F} where the 
authors suggest to expand the TDDFT response function 
$\chi_{\lambda}$ in a power series of the RPA response function 
(instead of the non-interacting one), and then to keep only the 
first order. This amounts in a alternative re-summation,
here named t$^{\rm \prime}$RPAx, for the interacting response function: 
\begin{equation}
 \chi^{\rm t^{\prime}RPAx}_{\lambda} = \chi_{\lambda}^{\rm RPA} + \lambda \chi_{\lambda}^{\rm RPA} f_{\rm x} \chi_{\lambda}^{\rm RPA}.
\label{eq3.5}
\end{equation}
\noindent  

We notice that tRPAx and t'RPAx both only require $h_{\rm x}$ to be defined. Both approximations  
thus neglect all higher order particle-hole scatterings which in the original RPAx are simulated by the kernel.

% (actually is the other way round: $f_x$ is defined once the first-order correction 
%to the response function $\chi_{\lambda}$ is given).

Up to first order the alternative RPAx response functions coincide 
with the original one,  while they have different power expansions
 starting from the $\lambda^2$ term, meaning that only contributions 
already approximated at the RPAx level are affected by these different
re-summations. 
%Finally we notice that tRPAx and t$^{\rm \prime}$RPAx
%can be obtained from Eq.~(\ref{eq1.4}) setting 
%$f_{\rm xc}^{\lambda} = \lambda f_{\rm x}/(1+\lambda \chi_0 f_{\rm x})$ 
%and $f_{\rm xc}^{\lambda} = \lambda f_{\rm x}/(1+\lambda \chi^{\rm RPA}_{\lambda} f_{\rm x})$ 
%respectively, explaining how the truncations mimic the correlation 
%contributions, that are missing at the bare RPAx level, by a 
%system-dependent renormalization of the bare $f_{\rm x}$.

Fig.~\ref{fig2.2} shows the correlation energies per particle obtained starting from the 
alternative RPAx approximations of the response function. As expected, for high density 
electron gases (small values of $r_s$) the correlation energies are essentially 
identical to the original one, since the underlying response functions are the same in 
the limit for $\lambda\rightarrow 0$. At the same time they are well behaved also where 
the original RPAx approximation breaks down. 

\begin{figure}[t]
\begin{center}
\includegraphics[scale=0.3]{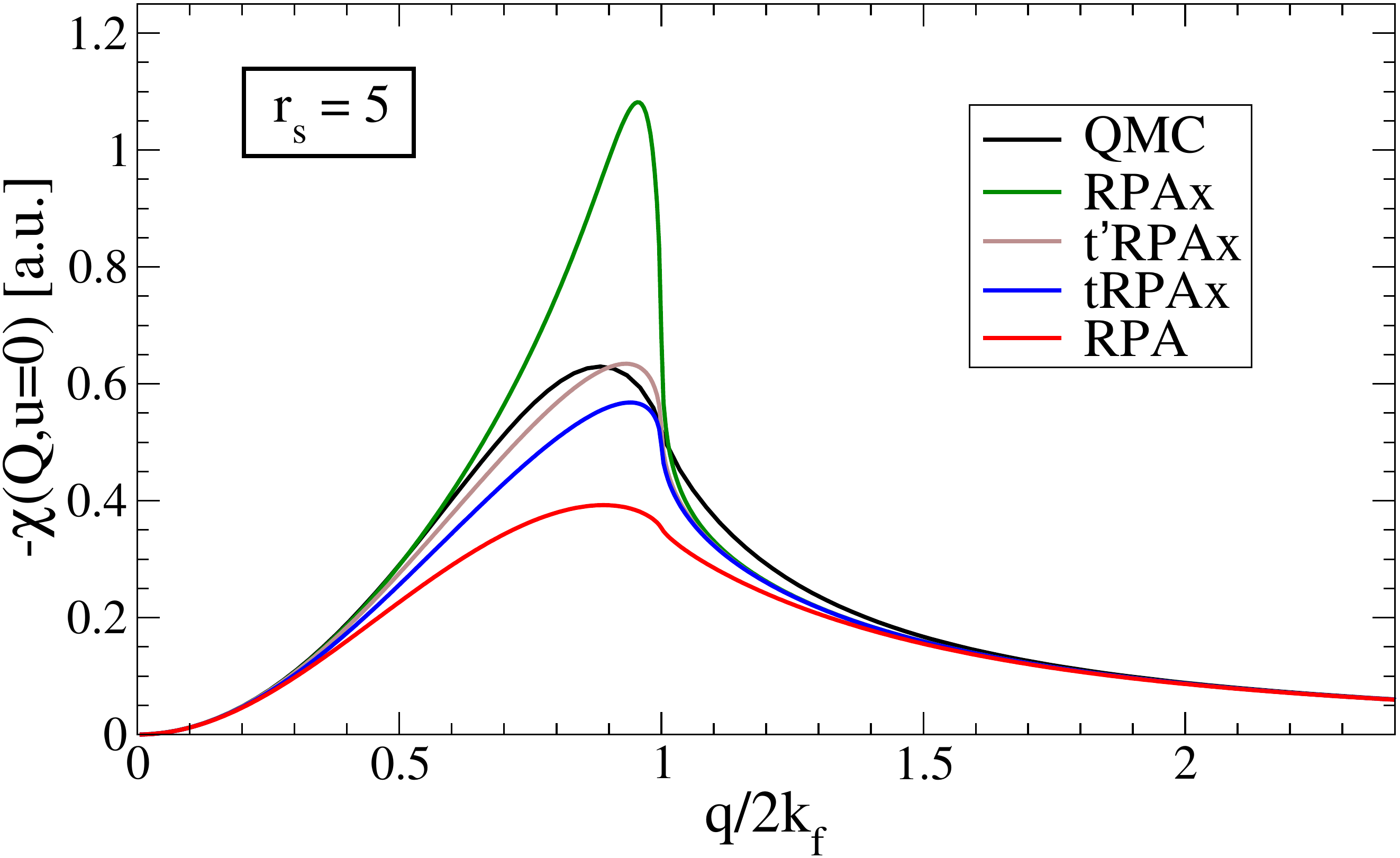}
\caption{Approximate static response functions for the 
HEG at $r_s=5$ as compared to the exact QMC calculations~\cite{moroni_static_1995}. 
The alternative RPAx approximations show a better agreement with QMC results.
%tRPAx and t$^{\rm \prime}$RPAx static response functions for different 
% values of $r_s$. No critical behavior is found even in the low density regime. 
}
\label{fig2.1}
\end{center}
\end{figure}

In Fig.~\ref{fig2.1} we compare the corresponding static density response functions 
(calculated at full interaction strength $\lambda=1$) with the exact one, obtained 
from QMC calculation~\cite{moroni_static_1995}, for a density 
corresponding to $r_s=5$. The difference between RPA and QMC
results reveals that exchange and correlation effects in the 
kernel are important already at this density; including the 
exact-exchange kernel (original RPAx) overcorrects the RPA
deficiency, in particular between $k_F$ and $2k_F$, while both the alternative RPAx approximations give
a much better agreement with accurate QMC calculations. 
Thus despite the fact that the RPAx energy is better at this value of 
$r_s$ the static response function is worse suggesting that the RPAx
results are subjected to a cancellation of errors when integrated over the frequency. 
%%%%%%%%%%%%%%%%%%%%%%%%%%%%%%%%%%%%%%%%%%%%%%%%%%%%%%%%%%%%%%%%%%%%%%%%%%%%%%%%%%%%%%%%%%%%%%%
%%%%%%%%%%%%%%%%%%%%%%%%%%%%%%%%%%%%%%%%%%%%%%%%%%%%%%%%%%%%%%%%%%%%%%%%%%%%%%%%%%%%%%%%%%%%%%%

%confirms the effectiveness of the alternative 
% re-summations in fixing the instability problem. At the static level
%the response functions of tRPAx and t$^{\rm \prime}$RPAx
%(evaluated at full interaction $\lambda=1$) do not show any critical behavior 
%in the range of density analyzed ($r_s$ up to 20); moreover when the density 
%decreases a trend opposite to the one found for the RPAx response function 
%(see Fig.~\ref{fig2}) is observed, suggesting no divergence would appear even for smaller 
%densities.

In the range of densities analyzed, tRPAx and t$^{\rm \prime}$RPAx response functions
do not show any critical behavior; moreover when the density decreases a trend opposite 
to the one found for the RPAx response function is observed with a reduction (instead of the 
enhancement shown in Fig.~\ref{fig2}) of the height of the peak near $2k_F$, suggesting 
no divergence would appear even for smaller densities.

\subsection{Spin-polarized HEG}\label {sect:sect3.3}

We continue our analysis of the HEG at the RPAx level by studying the spin 
magnetization dependence of the correlation energy of the system. 
We start noticing that for the non-interacting system the spin-up and 
spin-down components of the gas are independent so that a simple scaling 
relation between the non-interacting density-density response functions 
of the polarized and unpolarized gas can be derived:
\begin{align}
  \chi_0^{\uparrow \uparrow}[n^{\uparrow}] &= \frac{1}{2} \chi_0[2 n^{\uparrow}] \nonumber \\
  \chi_0^{\downarrow \downarrow}[n^{\downarrow}] &= \frac{1}{2} \chi_0[2 n^{\downarrow}]
\label{eq4.3}
\end{align}
while $\chi_0^{\uparrow \downarrow} = \chi_0^{\downarrow \uparrow} = 0$. 

The spin-up and spin-down components behave as independent constituents of 
the system at the exchange level too and a scaling relation similar to 
Eq~(\ref{eq4.3}) holds true also for the exchange energy~\cite{exchange-spin} and, accordingly, 
for the exchange potential and kernel:
\begin{equation}
 \begin{cases}
  \upsilon_{\rm x}^{\uparrow}[n^{\uparrow}]  = \upsilon_{\rm x}[2n^{\uparrow}]
%\left[\frac{\delta E_{\rm x}}{\delta n}\right]_{n=2n^{\uparrow}}
\\
  \upsilon_{\rm x}^{\downarrow}[n^{\downarrow}]  = \upsilon_{\rm x}[2n^{\downarrow}]
%\left[\frac{\delta E_{\rm x}}{\delta n}\right]_{n=2n^{\downarrow}}
\\
 \end{cases}
\quad
\begin{cases}
 f_{\rm x}^{\uparrow \uparrow}[n^{\uparrow}] = 2f_{\rm x}[2n^{\uparrow}]
%\frac{\delta^2 E_{\rm x}^p}{\delta n^{\uparrow} \delta n^{\uparrow}} = 2f_{\rm x}[2n^{\uparrow}] 
\\
 f_{\rm x}^{\downarrow \downarrow}[n^{\downarrow}] = 2f_{\rm x}[2n^{\downarrow}]
%\frac{\delta^2 E_{\rm x}^p}{\delta n^{\downarrow} \delta n^{\downarrow}} = 2f_{\rm x}[2n^{\downarrow}]
\end{cases}
%\quad\quad\quad f_{\rm x}^{\uparrow \downarrow} = f_{\rm x}^{\downarrow \uparrow} = 0
\label{eq4.5}
\end{equation}
while $f_{\rm x}^{\uparrow \downarrow} = f_{\rm x}^{\downarrow \uparrow} = 0$. Thus at 
the RPAx level the interaction between the spin-up and spin-down components 
of the system is only mediated by the Coulomb kernel $\upsilon_{\rm c}$.

Although more involved than for the unpolarized case, the solution of the 
RPAx Dyson equation for the polarized gas is anyway straightforward and 
using the definitions in Eq.~(\ref{eq4.3}) and Eq~(\ref{eq4.5}), the RPAx 
response function of the polarized HEG can be written as:
\begin{equation}
\chi_{\lambda} =  \frac{ \frac{1}{2} \left\{ \left[ \frac{\chi_0 }{1-\lambda \chi_0 f_{\rm x}} \right]_{2n^{\uparrow}} + 
                              \left[ \frac{\chi_0}{1-\lambda \chi_0 f_{\rm x}} \right]_{2n^{\downarrow}}\right\} } 
 {1 - \frac{1}{2} \left[ \frac{\lambda \chi_0 \upsilon_{\rm c}}{1-\lambda \chi_0 f_{\rm x}} \right]_{2n^{\uparrow}} 
    - \frac{1}{2} \left[ \frac{\lambda \chi_0 \upsilon_{\rm c}}{1-\lambda \chi_0 f_{\rm x}} \right]_{2n^{\downarrow}} }
\label{eq4.7}
\end{equation}
where $\chi_0$ and $f_{\rm x}$ are the same functions already used for the unpolarized 
case but evaluated at density $2n^{\uparrow}$ or $2n^{\downarrow}$.

\begin{figure}[t]
\begin{center}
\includegraphics[scale=0.3]{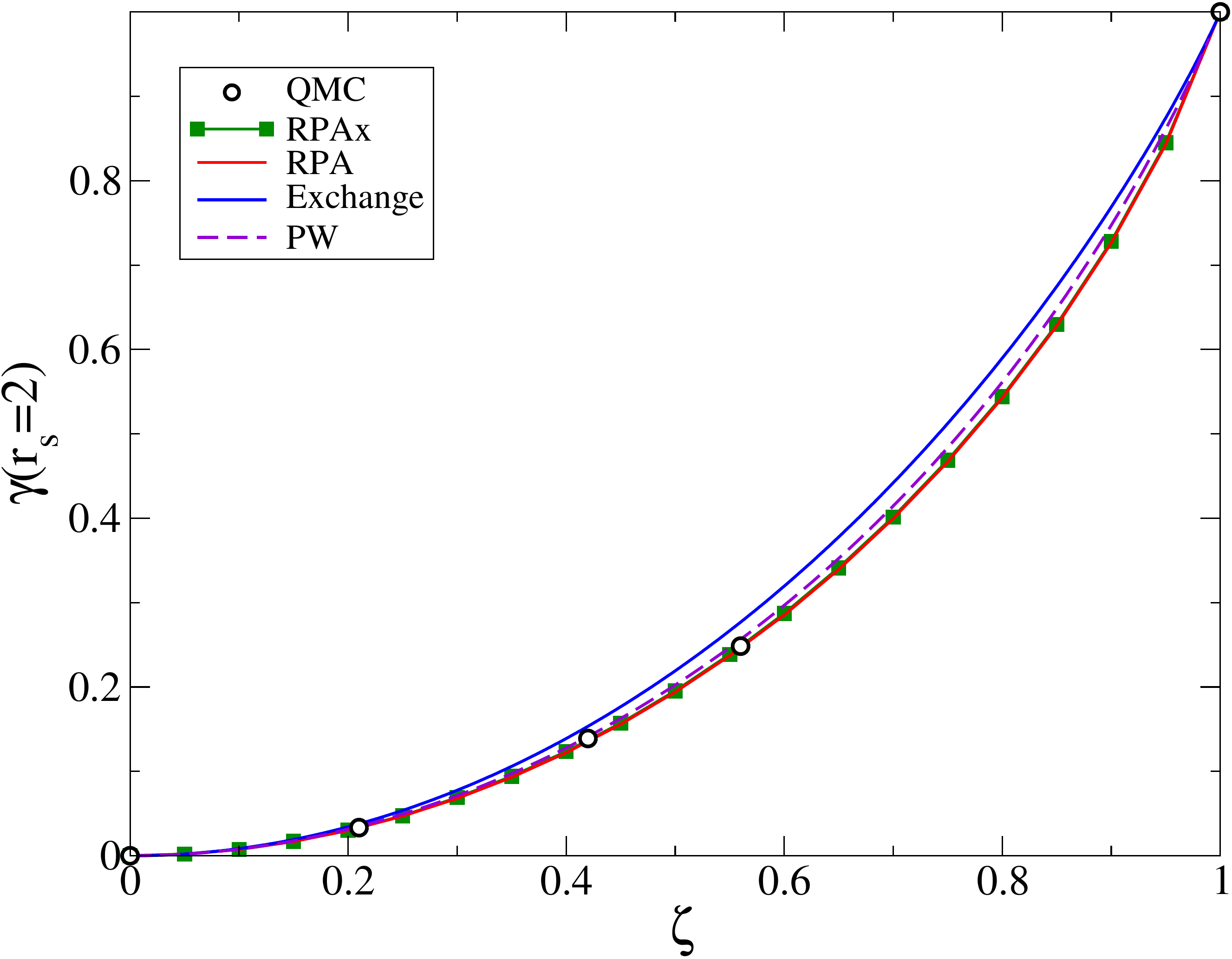}
\caption{Spin polarization function $\gamma$ for $r_s=2$ from RPA (red line), 
RPAX (green squares), Perdew-Zunger parametrization~\cite{Perdew-Zunger} 
(blue solid line), Perdew-Wang parametrization~\cite{Perdew-Wang} 
(brown dashed line) and Quantum Mote Carlo calculation~\cite{Ortiz}
(open circles).}
\label{fig3}
\end{center}
\end{figure}

Integrating Eq.~(\ref{eq1.3}) with the new definition of $\chi_{\lambda}$ 
in Eq.~(\ref{eq4.7}), gives the correlation energy per particle, $\epsilon_c$, 
as a function of $n^{\uparrow}$ and $n^{\downarrow}$ or, equivalently, 
as a function of $r_s$ and $\zeta$. 
At the RPA level the dependence of the correlation energy on the spin 
magnetization has been already calculated long time ago by Von Barth 
and Hedin~\cite{vBH} and more recently by Vosko, Wilk, and Nusair~\cite{VWN}. 
Our RPA results, simply obtained by setting $f_{\rm x}=0$ in Eq.~(\ref{eq4.7}), 
are, within the numerical accuracy, in perfect agreement with both the 
above mentioned calculations. Fig.~\ref{fig3} shows the spin-polarization 
function $\gamma$ defined as
\begin{equation}
 \gamma(r_s,\zeta) = \frac{\epsilon_c(r_s,\zeta) - \epsilon_c(r_s,0)}{\epsilon_c(r_s,1) - \epsilon_c(r_s,0)}
\label{eq4.9}
\end{equation}
for the case $r_s=2$ evaluated at the RPA and RPAx level, and compares it 
with the exchange-only dependence that is the one assumed in the Perdew-Zunger 
parametrization~\cite{Perdew-Zunger} of Local Spin Density Approximation (LSDA) 
and with the Perdew-Wang parametrization~\cite{Perdew-Wang}, that is based on 
the more physically motivated spin-interpolation expression proposed by Vosko, 
Wilk, and Nusair~\cite{VWN}.
While within RPAx the correlation energy significantly improves with respect 
to RPA results, there is essentially no difference between the RPA and RPAx 
spin-polarization functions. For this value of $r_s$, calculations 
done with the alternative re-summations (tRPAx and t$^{\rm \prime}$RPAx) 
give essentially the same results as the original RPAx and are not shown 
in Fig.~(\ref{fig3}). 
Thus for this property of the system RPA and all the RPAx (original and alternative) 
approximations give results in very good agreement with accurate Quantum Monte Carlo 
calculations\cite{Ortiz} performing much better than the Perdew-Zunger parametrization 
and slightly better than the more sophisticated Perdew-Wang parametrization.

\section{Bond dissociation of dimers}

As a second test for the RPAx approximation we studied the dissociation 
curve of the hydrogen and nitrogen molecules. 

\begin{figure}[t]
\begin{center}
\includegraphics[scale=0.3]{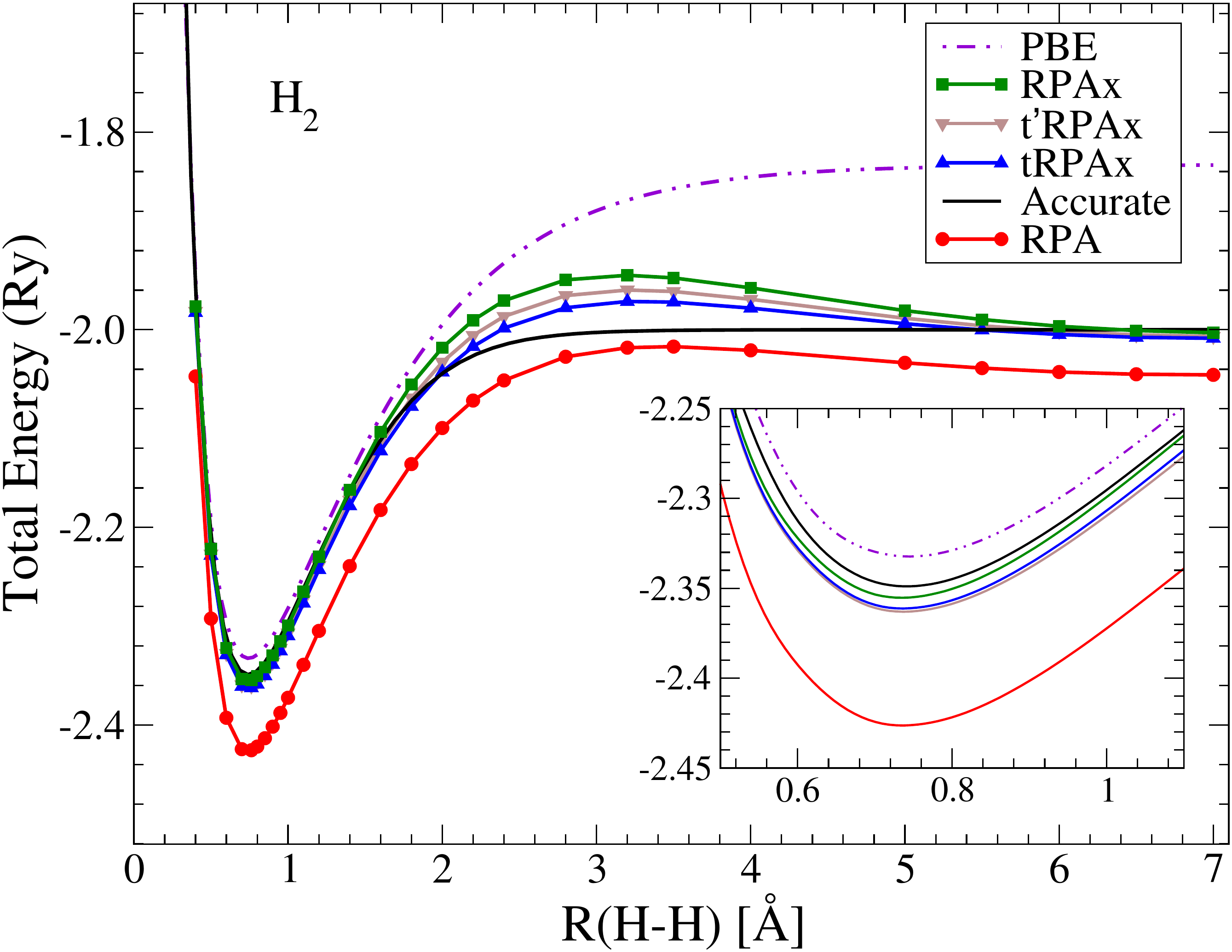}
\caption{Dissociation curve of H$_2$ molecule. PBE, RPA and RPAx (original and alternative) results
are compared with accurate calculations~\cite{kolos_potential_1965}. }
\label{fig:H2-diss}
\end{center}
\end{figure}

\begin{figure}[t]
\begin{center}
\includegraphics[scale=0.3]{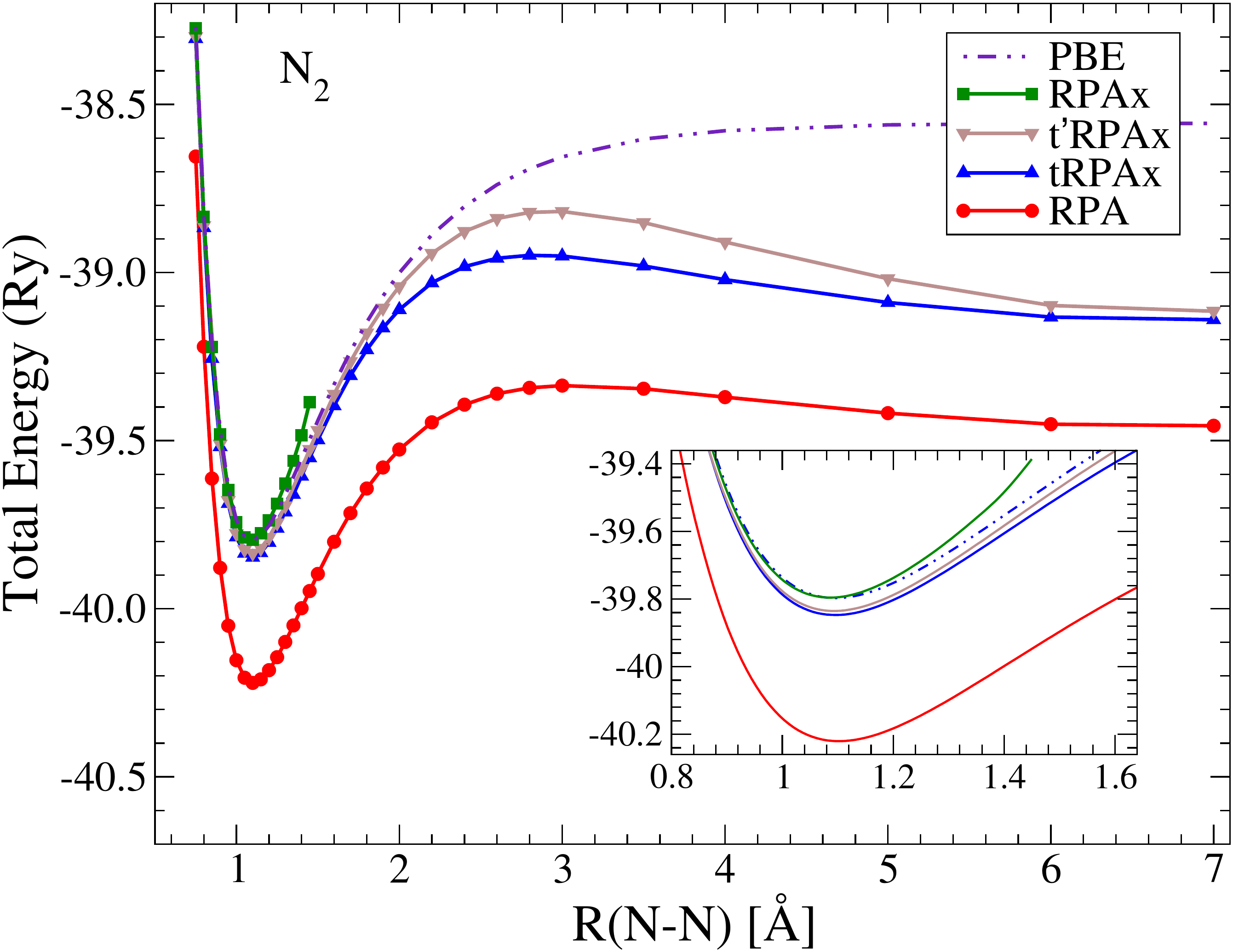}
\caption{Dissociation  curve of N$_2$ molecule. The plot compares results from PBE, RPA and
RPAx (original and alternative) calculations.}
\label{fig:N2-diss}
\end{center}
\end{figure}

%The dissociation of molecules into open-shell fragments represent the simplest and 
%paradigmatic example of situations where, from a quantum 
%chemistry perspective, multiple determinants associated with 
%degeneracy or near degeneracy is in principle needed and any 
%single determinant wave-function approach will fail. 
%H$_2$ and N$_2$ dissociation are two example of such a physical situation.
%Within standard Density Functional Approximations (DFAs) the proper 
%(singlet) KS ground state of these molecules at large interatomic 
%separations has too high total energy (as illustrated later 
%in Figs.~\ref{fig:H2-diss} and~\ref{fig:N2-diss}).
%reflecting the fact that the KS orbitals from standard DFAs are much too diffuse.
Within standard Density Functional Approximations (DFAs) the proper 
(singlet) KS ground state of these molecules at large interatomic 
separations has too high total energy (as illustrated later 
in Figs.~\ref{fig:H2-diss} and~\ref{fig:N2-diss}).
A better agreement with the experimental potential energy curve 
can be achieved resorting to a spin-polarized calculation which 
give good energies, however at the price of a qualitatively wrong 
spin-density. In a spin-unrestricted calculation, beyond a certain 
value of the interatomic separation the two spin components, 
defining the total electron density are no longer equal leading 
to solution which is no more a singlet as it should. 

The H$_2$ and N$_2$ dissociation curves at the RPA level has been previously 
studied~\cite{Fuchs,caruso_bond_2013,furche_molecular_2001}. 
In Refs.~\onlinecite{Fuchs,rohr} the authors have shown RPA to be size-consistent,
and thus to correctly describe the dissociation without resorting to any artificial 
spin-symmetry breaking. However the total energy is far too negative 
because of the well know overestimation of the correlation 
energy~\cite{RPA+}. Moreover an erroneous repulsion 
bump appears in the dissociation curves at intermediate distances.

Recently He\ss{}elmann \emph{et al.} reported the H$_2$ dissociation 
curve within the RPAx approximation showing very good result for the total energy both
around the equilibrium position $R_0$ and at dissociation but still 
the problem of the unphysical bump at intermediate bond-lengths 
remains.
G\"orling and coworkers have also computed RPAx total energies for a set of 21
molecules but always in their equilibrium geometries~\cite{hesselman_random_2010,bleiziffer_RI_2012}.

Here we would like to asses the performance of the RPAx 
(original and alternative) approximations for molecules beyond 
their equilibrium geometries studying the dissociation 
curves of H$_2$ and N$_2$.

The dimers and the corresponding isolated atoms were simulated 
using a simple-cubic super cell with a size length $a=22$ and $a=25$  
bohr, respectively. A kinetic energy cut-off of $50$ Ry were 
used for both systems and up to 200 lowest-lying eigenpairs of the 
generalized-eigenvalue problem in Eq.~(\ref{eq2.1.1}) were used
to compute the RPA and RPAx correlation energies. All the 
calculations have been done starting from well converged PBE orbitals.

%%%%%%%%%%%%%%%%%%%%%%%%%%%%%%%%%%%%%%%%%%%%%%%%%%%%%%%%%%%%%%%%%
\begin{table}[b]
\caption{\label{tab:struc-par} Equilibrium properties of hydrogen and nitrogen
dimers computed within different functionals: PBE, RPA and all the RPAx.  
Accurate values extracted from dissociation curves from Ref.~\onlinecite{kolos_potential_1965}
for H$_2$ and from Ref.~\onlinecite{gdanitz_accurately_1998} for N$_2$ are also given.
Equilibrium bond length ($R_0$) in \AA\, binding energy ($E_b$) in meV ,
and vibrational frequency ($\omega_0$) in cm$^{-1}$ }
\begin{ruledtabular}
\begin{tabular}{l l c c c c c c }
      & PBE & RPA & RPAx & tRPAx & t$^{\rm \prime}$RPAx & Refs. \\
\hline
H$_2$ & & & & & \\ 
      $\quad R_0$(\AA) 		  & 0.755 & 0.740 & 0.738 & 0.742 & 0.738 & 0.741 \\
      $\quad E_b$(eV) 		  & 6.78  & 4.85  & 4.41  & 4.48  & 4.45  &  4.75  \\
      $\quad \omega_0$(cm$^{-1}$) & 4219  & 4520  & 4560  & 4506  & 4406  &  4529 \\
\hline
N$_2$ & & & & & \\
      $\quad R_0$(\AA) 		  & 1.102 & 1.100 & 1.085 & 1.090 & 1.086 & 1.095 \\
      $\quad E_b$(eV) 		  & 16.86  & 9.92  & --  & 9.22  & 9.07  &  9.91  \\
      $\quad \omega_0$(cm$^{-1}$) & 2274  & 2322  & 2569  & 2430  & 2482  &  2383 \\
\end{tabular}
\end{ruledtabular}
\label{tab:struct-par}
\end {table}
%%%%%%%%%%%%%%%%%%%%%%%%%%%%%%%%%%%%%%%%%%%%%%%%%%%%%%%%%%%%%%%%%%%%%%%%%%%%%%%%%%%%%%%%%%%%%%%%
In Figs.~\ref{fig:H2-diss}  we report our results 
for the dissociation curves of H$_2$ and in 
Tab~\ref{tab:struct-par} the structural parameters extracted from them.
Comparison with accurate calculations~\cite{kolos_potential_1965}
illustrates the aforementioned deficiencies of PBE and RPA dissociation curves:
standard DFAs give too high total energy in the dissociation limit while RPA 
overestimates the correlation energy leading to a curve
well below the reference one.
%~\footnote{
%For the N$_2$ case we do not have an accurate reference to compare with 
%since we are using a pseudo-potential approach. Nevertheless we can assume that 
%an hypotetical accurate calculation would lie, at equilibrium geometry, near 
%the PBE calculation, which is known to give a good description of molecules
%near the minimum~(CITAZIONE)
%}. 
Including the exact-exchange kernel leads to a sensible improvement
in the total energy description; as can be seen from the inset 
in Fig.~\ref{fig:H2-diss} the RPAx total energies around the 
equilibrium position is in very good agreement with accurate 
quantum chemistry calculations. The alternative re-summations
while essentially giving the same energy as the original RPAx 
in the minimum region, have a positive effect on the dissociation 
curve at intermediate distances reducing the height of the repulsive 
hump. We notice that at large interatomic separations all 
the RPAx approximation drops below the exact dissociation limit of $2$ Ry
in agreement with the analysis reported in Ref.~\onlinecite{Fuchs}.

With the simple H$_2$ example in mind we can turn to analyze 
the more interesting case of the N$_2$ molecule.  
In Fig.~\ref{fig:N2-diss} we report our results for the dissociation curve and 
in Tab.~\ref{tab:struct-par} the structural parameters obtained from them. 
As already observed for the H dimer also in this case the whole RPA 
dissociation curve lies far below all the other curves. Nevertheless
the structural parameters at the RPA level are in very good agreement 
with results from accurate quantum chemistry calculations~\cite{gdanitz_accurately_1998}.
Including the exact-exchange contribution to the kernel 
corrects for the RPA overestimation of the correlation energy 
shifting the RPAx dissociation curve upward. At the same time, 
the good performance for the equilibrium bond length and 
the vibrational frequency already obtained at the RPA level
is maintained.
However, unlike what happens for the H$_2$ molecule, in this case the original RPAx 
approximation breaks-down when the nitrogen atoms are separated. 
For bond lengths greater than $R=1.45$\AA\ the RPAx response function 
is no more negative-definite leading to an instability which is
very similar the one observed for the low-density homogeneous 
electron gas and, ultimately, causes the break-down of the approximation.
%for the N$_2$ molecule in its stretched geometry. 
The alternative re-summations proposed to fix the pathological behavior 
of the RPAx response function in the HEG, turn out to be effective also in this very 
different situation. The tRPAx and t$^{\rm \prime}$RPAx dissociation curves are close to the 
RPAx one in the equilibrium region (see the inset in Fig.~\ref{fig:N2-diss}) 
but they are well-behaved also for bond lengths greater that $R=1.45$\AA\ 
overcoming, also in this case, what appears to be an intrinsic inadequacy 
of the original RPAx approximation. 

%The tRPAx and t$^{\rm \prime}$RPAx dissociation curves are very similar to the 
%RPAx one in the equilibrium region (see inset in Fig.~\ref{fig:N2-diss}) 
%but they are well-behaved also for bond lengths greater that $R=1.45$\AA\ 

%Therefore tRPAx and t$^{\rm \prime}$RPAx are able to cure 
%what appears to be an intrinsic inadeguacy of the original 
%RPAx approximation. 

\section{Conclusions}\label {sect:sect5}

In this work, we have setted the RPAx approximation for the correlation energy
within a general scheme that combines the general framework of the ACFD theory 
with a systematic many-body approach along the adiabatic-connection path
and allows in principle to improve the xc kernel for the purpose of 
calculating increasingly more accurate correlation energy. We have 
shown that, in a perturbative approach, RPA is an ``incomplete'' 
approximation and that the exact-exchange kernel has to be taken into account 
for a consistent description to first order in the interaction strength.
An efficient method for the calculation of the RPAx correlation energy has been 
proposed, based on an eigenvalue decomposition of the time-dependent response 
function of the many body system in the limit of  vanishing coupling constant.

The accuracy of the RPAx approximation has been tested on the homogeneous electron 
gas revealing a great improvement over RPA results and a very good 
agreement with accurate QMC calculations. The spin magnetization dependency of 
the RPA and RPAx correlation energies has been calculated as well, showing a big 
improvement if compared to standard parametrization and a nearly perfect 
agreement with QMC calculation.

These encouraging results are however disturbed by the break-down of the 
procedure for large values of $r_s$ where the RPAx density-density response 
function unphysically changes sign thus indicating that correlation contributions 
to the kernel are needed to obtain accurate results for the HEG at low densities. 
Staying within an exact first order approximation to the particle-hole
interaction we have suggested two simple and inexpensive modifications of the RPAx 
approximation which lead to a good description of the correlation energy of the system even 
in the limit of small densities. 

We then examine molecular dissociation of H$_2$ and N$_2$ within the RPAx approximation, discovering 
the same virtues and vices already observed in the HEG case. A sensible improvement 
of the total energy description is disturbed by a pathological behavior of the response
function which ultimately poses doubts on the broad applicability of the RPAx approximation.
The alternative re-summations, tRPAx and t$^{\rm \prime}$RPAx, proposed here,
have been shown to be able to fix the RPAx inadequacy without compromising its virtues.  
Although more tests are needed in order to completely characterize them, tRPAx and t$^{\rm \prime}$RPAx 
emerge as promising and stable alternatives to the original RPAx approximation.

\begin{acknowledgments}
We thank CINECA (Bologna, Italy) and SISSA (Trieste, Italy) for the availability 
of high performance computing resources and support. 
\end{acknowledgments}

\end{document}